# Superconductivity in centrosymmetric topological superconductor candidate TaC

D. Y. Yan[1,2], M. Yang[1,2], C. X. Wang[1,2], P. B. Song[1,2], C. J. Yi[1,*], Y. G. Shi[1,3,*]

*[1] Beijing National Laboratory for Condensed Matter Physics and Institute of Physics, Chinese Academy of Sciences, Beijing 100190, China*

*[2] School of Physical Sciences, University of Chinese Academy of Sciences, Beijing 100190, China*

*[3] Center of Materials Science and Optoelectronics Engineering, University of Chinese Academy of Sciences, Beijing 100049, China*

\* To whom correspondence should be addressed.

E-mail: ygshi@iphy.ac.cn, chjyi@iphy.ac.cn

**Abstract**

We report the synthesis and physical properties of the single crystals of TaC, which are proposed to hold topological band structure as a topological superconductor (TSC) candidate. Magnetization, resistivity and specific heat measurements are performed and indicate that TaC is bulk superconductor with critical temperature of 10.3 K. TaC is a strongly coupled type-II superconductor and the superconducting state can be well described by *s*-wave Bardeen-Cooper-Schrieffer (BCS) theory with a single gap. The upper critical field ($H_{c2}$) of TaC shows linear temperature dependence, which is quite different from most conventional superconductors and isostructural NbC, which is proposed to manifest topological nodal-loops or type-II Dirac points as well as superconductivity. Our results suggest that TaC would be a new candidate for further research of TSCs.

**Introduction**

Exploration of topological superconductors (TSCs) is progressively becoming a fruitful subject in condensed matter physics. A TSC has topologically nontrivial superconducting phases, characterized by a full superconducting gap in the bulk, but supporting a protected Majorana zero-energy mode (MZM) at the vortex core, as well as gapless edge or surface states [1-3]. Because the Majorana bound state obeys non-Abelian statistics and is theoretically predicted to emerge in certain *p*-wave TSCs, it provides a possible way to approach fault-tolerant quantum computation [4]. The topological surface state and MZM have been found to exist in single material platforms of FeTe$_{0.55}$Se$_{0.45}$ bulk single crystals [5-8] and similar compounds of iron-based superconductors [9-11]. In a TSC, it is the wave function of the electron pairs that exhibits topological properties, and among the most promising candidates for topological superconductors are materials with noncentrosymmetric (NCS) crystal structures. NCS crystal structure with broken inversion symmetry has antisymmetric spin-orbit coupling (SOC) and spin degeneracy, which gives rise to topological phases and allowing the mixing of spin-triplet and spin-singlet pairing channels [12-14]. As yet, the effort to search for TSCs continues, but few topological NCS superconductors have been identified. In recent years, the observation of topological surface states (TSSs) in some centrosymmetric superconducting materials has attracted most interest. For centrosymmetric superconductors, the parity mixing of pair potential may occur near the surface where the inversion symmetry is broken. Hence, a centrosymmetric superconductor may become a great candidate for TSC [15]. It has been found that the centrosymmetric $\beta$-PdBi$_2$ has TSSs at $E_F$ in the normal state and fully gapped superconductivity in the bulk, which makes it a great candidate for achieving TSC states [16-21]. Meanwhile, the anisotropic Majorana bound states were theoretically predicted and experimentally observed in 2$M$-WS$_2$, which is a centrosymmetric superconductor with $T_c \sim 8.8$ K. [22,23]. SnTaS$_2$ has also been identified to be a superconductor with a possible topological nodal line semimetal character [24]. Noteworthily, Z. H. Cui *et al.* proposed that type-II Dirac semimetal states exist in the band structure of TaC [25]. T. Shiroka *et al.* reported the superconducting properties of the polycrystalline NbC and TaC and calculated their topological band structures, pointing out the potential TCSs states among these compounds [26]. Meanwhile, combining the theoretical calculation and angle resolved photoemission spectroscopy (ARPES) measurements, we observe that NbC is a type-II Dirac semimetal with robust Fermi-surface nesting that is responsible to the strong electron-phonon interaction [27].

In this work, we successfully grow the single crystals of TaC and report the magnetism, resistivity and specific heat properties. The results indicate that TaC is BCS superconductor with *s*-wave single gap in bulk state. However, the electron-phonon coupling (EPC) strength of TaC is larger than most conventional BCS superconductors. In addition, the upper critical field shows anomalous linear temperature dependence at low temperature, which is usually present in some iron-based superconductors.

## Experiment details

Single crystals of TaC were grown by using Co-flux mothed. Starting materials of high-purity Ta, C and Co were mixed and loaded in an alumina crucible at a molar radio of Ta : C : Co = 1 : 1 : 9. The alumina crucible was then placed in an argon-filled furnace and heated to 1500 °C. After a dwell time of 20 hours, the crucible was slowly cooled to 1300 °C at a rate of 1 °C/h and then cooled naturally down to room temperature. Finally, the excess Co was removed by immersion in nitrohydrochloric acid for one day, yielding golden TaC single crystals in rectangle shape.

To investigate the crystalline structure, single-crystal x-ray diffraction (XRD) was carried out on Bruker D8 Venture diffractometer at 293 K using Mo $K\alpha$ radiation ($\lambda$ = 0.71073 Å). The crystalline structure was refined by full-matrix least-squares method on $F^2$ by using the SHELXL-2016/6 program. Selected crystals were used for magnetic susceptibility ($\chi$), electrical resistivity ($\rho$) and specific heat ($C_p$) measurements. The magnetic properties were measured in a Magnetic Properties Measurement System (MPMS-III, Quantum Design Inc.) under fixed applied magnetic field of 20 Oe in field-cooling (FC) and zero-field-cooling (ZFC) modes. Isothermal magnetization (*M-H*) was measured at several fixed temperatures by sweeping applied field. The electrical resistivity and specific heat were measured in a Physical Property Measurement System (PPMS, Quantum Design Inc.) by using a standard dc four-probe technique and a thermal relaxation method, respectively.

## Result and discussion

The single-crystal XRD study reveals that TaC crystallize in a centrosymmetric NaCl-type structure with space group of *Fm-3m* (No. 225). The detailed crystallographic parameters are summarized in Table I. Figure 1(a) shows the photograph of TaC crystal, and the back square of 1×1 mm indicates the size of the crystal. A schematic drawing of the crystal structure is shown in Fig. 1(b). The XRD patterns of a flat surface of the single crystal is presented in Fig. 1(c), where only the *h00* peaks are detected. And the Laue diffraction pattern with fourfold symmetry on the (100) surface of TaC is presented in the inset of Fig. 1(c), further proving the quality of the single crystal.

Temperature dependent magnetic susceptibility ($4\pi\chi$) with an applied field $H = 20$ Oe was measured from 2 K to 16 K, as shown in Fig. 2 (a). The critical temperature $T_c$ was determined by the intersection between the normal state of $4\pi\chi$ extrapolated to lower temperature and the superconducting state corresponding to the steepest slope of the diamagnetic area [28], yielding that $T_c$ 10.3 K for TaC. The superconducting volume fraction calculated from the original magnetic measurement exceeds 150%, which comes from the demagnetization field due to the sample. Using the following expression [29]:

$$\chi = \frac{\chi_{exp}}{1-N\chi_{exp}} \quad (1)$$

where $N = 0.27$ is the demagnetizing factor estimated for the cuboid sample [30], the modified superconducting volume fraction is closed to 100%, as shown in Fig. 2 (a). Figure 2 (b) shows the *M-H* loops in various temperatures for TaC, indicating that TaC is type-II superconductors [31].

The temperature dependence of resistivity measured from 300 to 2 K is shown in Fig. 3 (a) for TaC. The resistivity undergoes a sudden reduction near $T_c$. The onset and zero-resistance temperature were marked in Fig. 3 (a). The zero-resistance temperature is consistent with the $T_c$ measured from magnetic susceptibility. Above $T_c$, the resistivity metallically goes down with cooling temperature, and the data can be well described by Bloch-Grüneisen law [32-35] with a formula of:

$$\rho(T) = \rho_0 + \alpha\left(\frac{T}{\theta_D}\right)^5 \int_0^{\theta_D/T} \frac{x^5}{(e^x-1)(1-e^{-x})}dx \quad (2)$$

where $\rho_0$ is the residual resistivity of the nomal state, $\alpha$ is proportional to electron-phonon coupling constant and $\theta_D$ is Debye temperature. The fit is shown as a red solid line in Fig. 3 (a), giving out the Debye temperature $\theta_D = 213.5$ K. The Fig. 3 (b) shows the magnetic field dependent resistivity at various fixed applied magnetic fields, indicating that the superconductivity is suppressed when magnetic field is applied. It should be noticed that the electrical resistivity shows multiple transitions by increasing the magnetic field. The second transition may be caused by the carbon-deficient in TaC single crystal, as previously reported that the superconducting transition temperature decreases with decreasing C content in TaC$_x$ (0 < x < 1) [36, 37].

With fitted parameters of $\theta_D$ and $T_c$, the EPC constant $\lambda_{ep}$ can then be calculated by using the inverted McMillan equation [38]:

$$\lambda_{ep} = \frac{1.04+\mu^*\ln(\frac{\theta_D}{1.45T_c})}{(1-0.62\mu^*)\ln(\frac{\theta_D}{1.45T_c})-1.04} \quad (3)$$

where the $\mu^*$ represents the repulsive screened Coulomb part, which is set to 0.13 for transition metal element. This yields a value of the superconducting parameter $\lambda_{ep} = 0.986$. Typically, materials with $\lambda_{ep} \rightarrow 1$ are classified as strongly coupled superconductors, while $\lambda_{ep} \rightarrow 0.5$ indicates weak coupling [39]. The relatively large $\lambda_{ep}$ indicates strong EPC strength in TaC. The large EPC constant was also observed in the isostructural compounds NbC [27] and NbC$_{1-x}$N$_x$ [40], which may be caused by the existence of Fermi-surface nesting at the fermi level, leading to the emergence of Kohn anomaly and enhance the EPC.

The specific heat data in various applied fields are given in Fig. 4 (a) for TaC. With increasing fields, the superconducting peak is gradually suppressed to lower temperature, which is consist with the behavior in resistivity as aforementioned. Fig. 4 (b) plots $\Delta C_P = C(0T) - C(0.5T)$ versus $T$. Under the applied magnetic field, the jump of $C_P$ due to superconducting transition is totally suppressed. Since phonon and normal state electronic specific heat are field independent, $\Delta C_P$ is the specific heat after subtracting the background originating from the lattice and normal state electronic contribution, yielding $\Delta C_P = C_{es} - \gamma_n T$, where $C_{es}$ is the quasiparticle contribution and $\gamma_n$ here means the normal state Sommerfeld coefficient of the superconducting part.

We fit $\Delta C_P$ with different gap functions as done in refs [41-43]. In BCS theory, the entropy $S_{es}$ in the superconducting state is written as:

$$S_{es} = -\frac{3\gamma_n}{k_B\pi^3}\int_0^{2\pi}\int_0^{\infty}[(1-f)\ln(1-f)+f\ln f]\,d\varepsilon d\phi \quad (4)$$

where $k_B$ is the Boltzmann constant and $f$ stands for the quasiparticle occupation number $f = (1+e^{E/k_BT})^{-1}$ with $E = \sqrt{\varepsilon^2 + \Delta^2(\phi)}$. $\Delta(\phi)$ is the angle dependence of the gap function. For a conventional $s$-wave superconductor $\Delta(\phi) = \alpha\Delta_{BCS}(T)$, which is angle independent. While for a standard $d$-wave superconductor

$\Delta(\phi) = r\Delta_{BCS}(T) \cos 2\phi$. Here $\Delta_{BCS}(T)$ is the weak coupling BCS gap function, and $r$ is a dimensionless factor. The electronic specific heat is calculated by $C_{es} = T(\partial S/\partial T)$. To allow for the possibility of any part of the sample being nonsuperconducting, we set $\gamma_n$ to vary in the fit, so the free parameters are $r$, $\gamma_n$ and $T_c$. As seen in Fig. 4 (b), the *s*-wave model well reproduces the experimental data, while the *d*-wave model deviates significantly from the data at low temperature. This result is same as NbC [27]. From the *s*-wave fit results, we get the parameters $r = 1.17$, $\gamma_n = 2.21$ mJ mol$^{-1}$ K$^{-2}$ and $T_c = 9.7$ K for TaC. It should be note that the $T_c$ of 9.7 K gotten from specific heat data is smaller than the data 10.3 K measured from magnetic and resistivity data for TaC. Consequently, the estimated superconducting gap $\Delta(0)$ is 1.72 meV. The entropy-conserving construction at $T_c$ gives $\Delta C/\gamma_n T_c = 1.94$, which obviously deviates from the weak coupling BCS value 1.43. And the value of $2\Delta(0)/k_B T_c$ is 4.12, which is also slightly larger than the BCS theory value 3.52.

The phase diagram of the upper critical field $H_{c2}(T)$ and the lower critical field $H_{c1}(T)$ versus $T$ are plotted, as shown in Fig. 5 (a). $H_{c1}$ was defined from the point deviating from the *M-H* linear curve due to the Meissner effect. As shown in the inset of Fig. 5 (a), the $H_{c1}(T)$ were fitted by the empirical power law expression $\mu_0 H_{c1}(T) = \mu_0 H_{c1}(0)\left[1 - \left(\frac{T}{T_c}\right)^b\right]$ with fixed $T_c = 10.3$ K. The fits reveal that $\mu_0 H_{c1}(0) = 38.6$ mT. The upper critical field $H_{c2}$ was defined from the 90% and 50% of the normal state resistivity as well as the zero point of the resistivity drop in Fig. 3 (b) for every magnetic field. Furthermore, the upper critical field was determined from the *M-H* curves in Fig. 2 (b) and the specific heat data in Fig. 4 (a). One can see that the $H_{c2}$ data of 90% and 50% $\rho$ very strangely deviate from other data, which needs further detailed studies for explanation. The $H_{c2}(T)$ curve of TaC determined from zero resistivity, magnetization and specific heat data shows an almost linear *T*-dependence at low temperature, which cannot be fitted by WHH model [44,45]. This anomalous behavior is different from most conventional superconductors [31,44] and NbC single crystals reported by us in the ref [27], as shown in Fig 5. (b). The linear *T*-dependence of $H_{c2}(T)$ is similar to some iron-based superconductors with multiband superconductivity [46,47]. Considering the second transition in the resistivity measurement under magnetic field, the linear $H_{c2}(T)$ may also be caused by the existence of the carbon-deficient in TaC. By linear extrapolations, the upper critical field is estimated to be 0.3 T, which is obviously smaller than the Pauli paramagnetic limit of 19 T, indicating that the orbital pair breaking is the essential mechanism and limits the upper critical field in TaC.

In addition, other parameters related to superconductivity can be calculated. By using $\lambda_{ep}$, $\gamma_n$, and $k_B$, the density of electronic states at the Fermi energy $N(E_F)$ can be calculated from the formula [48]:

$$N(E_F) = \frac{3\gamma_n}{\pi^2 k_B^2 (1+\lambda_{ep})} \tag{5}$$

Thus the $N(E_F)$ was estimated to be 0.69 eV$^{-1}$ per formula unit (f.u.). The mean free path $l$ can be determined by using following equation [49]:

$$l = 2.372 \times 10^{-14} \frac{(\frac{m^*}{m_e})^2 V_M^2}{N(E_F)^2 \rho_0} \tag{6}$$

where $V_M$ is the molar volume. Assuming that $\frac{m^*}{m_e} = 1$, the $l$ was calculated to be 448 Å.

The determined $H_{c2}(0)$ value can be used to calculate the Ginzburg-Landau coherence length $\xi_{GL}$ by using the equation [31]:

$$\mu_0 H_{c2}(0) = \frac{\Phi_0}{2\pi \xi_{GL}^2} \tag{7}$$

where $\Phi_0$ is the quantum flux $\frac{h}{2e}$. The value of $\xi_{GL}$ for TaC was calculated to be 331 Å.

Using the result of $\xi_{GL}$ with $H_{c1}(0)$, the Ginzburg-Landau penetration depth $\lambda_{GL} = 595$ Å of TaC were estimated by the lower critical field equation [50]:

$$\mu_0 H_{c1}(0) = \frac{\Phi_0}{4\pi \lambda_{GL}^2} \ln(\frac{\lambda_{GL}}{\xi_{GL}} + 0.49693) \tag{8}$$

From the Ginzburg-Landau penetration depth and the coherence length, the Ginzburg-Landau parameter is $\kappa_{GL} = \frac{\lambda_{GL}}{\xi_{GL}} > \frac{1}{\sqrt{2}}$, which again confirms that TaC is type-II superconductors [31]. Combining the results of $H_{c1}(0)$, $H_{c2}(0)$ and $\kappa_{GL}$, the thermodynamic critical field $H_c(0)$ was estimated to be 136 mT from the equation

[31]:

$$H_{c1}H_{c2} = H_c^2\ln(\kappa_{GL} + 0.08) \qquad (9)$$

The relationship between the BCS coherence length $\xi_0$ and the Ginzburg-Landau coherence $\xi_{GL}$ at $T = 0$ K is [51]

$$\frac{\xi_{GL}(0)}{\xi_0} = \frac{\pi}{2\sqrt{3}}(1 + \frac{\xi_0}{l})^{-0.5} \qquad (10)$$

From the above equation, we get the value of $\xi_0$ is 543 Å. The value of $\xi_0/l$ is close to 1, indicating that TaC superconductor is on the boundary between the dirty and clean limits. In summary, the observed and estimated superconducting parameters of TaC and NbC are listed in Table III for comparison.

Superconductors with nontrivial band structure provide possibility of realizing topological superconductors. Superconductivity can be combined with novel band structures by charge carrier doping or heterostructures constructed by superconductors and topological insulators. For instance, topological insulator $Bi_2Se_3$ becomes superconductor when intercalated by Cu, Sr, or Nb atoms [52-54]. And the heterostructure of $NbSe_2/Bi_2(Se, Te)_3$ is reported to hold TSC states at its interface [55,56]. However, doping can induce disorder and inhomogeneity effect, which hinder the unambiguous clarification of superconducting states in doped topological insulators. The difficulty of fabricating heterostructure and observing the interface-related phenomena also limits further studies of TSC. Different from the two strategies mentioned above, the coexistence of the bulk superconductivity and topological band structures in the same compound might generate intrinsic TSC with nontrivial surface states. For example, nontrivial topological band structures have been observed in $\beta$-$PdBi_2$ [16] and $PbTaSe_2$ [57] superconductors, which provide excellent playground for studying TSC. However, their superconducting temperature are relatively low (< 5 K), limiting the deeper investigations and possible applications.

Our experiments prove that TaC is a BCS superconductor with single *s*-wave gap in bulk. It has simple structure and high resistance to corrosion with high $T_c$ of 10.3 K. Meanwhile, TaC shows some unusual superconductivities such as strong EPC constant and linear temperature dependence of $H_{c2}$, which is distinct from most conventional superconductors as well as the isostructural NbC single crystal we reported recently [27]. The theoretical studies and ARPES measurements for NbC confirm that the superconductivity as well as the topological band structures could emerge together in this compound. The Fermi-surface nesting was observed in NbC by APPES, which is corresponding to the strong EPC. However, the ARPES research for TaC is still absent due to the difficulties on cleavage. First-principles calculation of TaC by Z. H. Cui *et al.* indicates the type-II Dirac point near Γ point in the Brillouin zone [25]. They also claim that Ta *d* orbitals and C *p* orbitals consist the gapless Dirac point with nontrivial topology. Therefore, the successful synthesis of single crystals of TaC may enlarge the measureable materials family of TSCs and provide new opportunity for further research on novel physical properties.

## Conclusion

In summary, we present the studies of magnetism, resistivity and specific heat on single crystal TaC, which is characterized as a type-II superconductor with centrosymmetric structure. The electronic specific heat in the superconducting state is well described by the single gap *s*-wave BCS expression. Most of superconducting properties for TaC are similar to NbC single crystals we reported in ref. [27]. The value of electron-phonon coupling constant is quite larger compared with conventional BCS superconductors, which may be caused by the Fermi-surface nesting in its electronic structures. Meanwhile, the linear behavior of $H_{c2}$ vs *T* for TaC is different from NbC and most BCS superconductors, making TaC a valuable material for further study. Hence, we suggest that TaC is a potential candidate of TSCs for further investigating the novel physical properties.

## Acknowledgements

We acknowledge Y. Wei, W. S. Hong for the assistance during the measurements and Z. L. Feng, X. Wang, W. S. Wei, F. F. Zhu for useful discussions. This work was supported by the National Key Research and Development Program of China (Grants No. 2017YFA0302901 and No. 2016YFA0300604), the National Natural Science Foundation of China (Grants No. 11774399), Beijing Natural Science Foundation (Grants No. Z180008), the K. C. Wong Education Foundation (Grants No. GJTD-2018-01) and the China Postdoctoral Science Foundation (Grants No. 2020M670504).

# Reference

[1] Teo, Jeffrey CY, and Charles L. Kane. "Majorana fermions and non-Abelian statistics in three dimensions." Physical review letters 104.4 (2010): 046401.

[2] Qi, Xiao-Liang, et al. "Time-reversal-invariant topological superconductors and superfluids in two and three dimensions." Physical review letters 102.18 (2009): 187001.

[3] Deng, Shusa, Lorenza Viola, and Gerardo Ortiz. "Majorana modes in time-reversal invariant s-wave topological superconductors." Physical Review Letters 108.3 (2012): 036803.

[4] Nayak, Chetan, et al. "Non-Abelian anyons and topological quantum computation." Reviews of Modern Physics 80.3 (2008): 1083.

[5] Yin, J. X., et al. "Observation of a robust zero-energy bound state in iron-based superconductor Fe(Te,Se)." Nature Physics 11.7 (2015): 543-546.

[6] Xu, Gang, et al. "Topological superconductivity on the surface of Fe-based superconductors." Physical review letters 117.4 (2016): 047001.

[7] Zhang, Peng, et al. "Observation of topological superconductivity on the surface of an iron-based superconductor." Science 360.6385 (2018): 182-186.

[8] Wang, Dongfei, et al. "Evidence for Majorana bound states in an iron-based superconductor." Science 362.6412 (2018): 333-335.

[9] Zhang, Peng, et al. "Multiple topological states in iron-based superconductors." Nature Physics 15.1 (2019): 41-47.

[10] Shi, Xun, et al. "FeTe$_{1-x}$Se$_x$ monolayer films: towards the realization of high-temperature connate topological superconductivity." Science bulletin 62.7 (2017): 503-507.

[11] Liu, Qin, et al. "Robust and clean Majorana zero mode in the vortex core of high-temperature superconductor (Li$_{0.84}$Fe$_{0.16}$)OHFeSe." Physical Review X 8.4 (2018): 041056.

[12] Bauer, Ernst, and Manfred Sigrist, eds. Non-centrosymmetric superconductors: introduction and overview. Vol. 847. Springer Science & Business Media, 2012.

[13] Khan, Mojammel A., et al. "Complex superconductivity in the noncentrosymmetric compound $Re_6Zr$." Physical Review B 94.14 (2016): 144515.

[14] Kneidinger, F., et al. "Superconductivity in non-centrosymmetric materials." Physica C: Superconductivity and its Applications 514 (2015): 388-398.

[15] Guan, Syu-You, et al. "Superconducting topological surface states in the noncentrosymmetric bulk superconductor $PbTaSe_2$." Science advances 2.11 (2016): e1600894.

[16] Sakano, M., et al. "Topologically protected surface states in a centrosymmetric superconductor β-$PdBi_2$." Nature communications 6.1 (2015): 1-7.

[17] Iwaya, Katsuya, et al. "Superconducting states of topological surface states in β-$PdBi_2$ investigated by STM/STS." APS 2016 (2016): E28-011.

[18] Herrera, E., et al. "Magnetic field dependence of the density of states in the multiband superconductor β−$Bi_2Pd$." Physical Review B 92.5 (2015): 054507.

[19] Kačmarčík, J., et al. "Single-gap superconductivity in β−$Bi_2Pd$." Physical Review B 93.14 (2016): 144502.

[20] Lv, Yan-Feng, et al. "Experimental signature of topological superconductivity and Majorana zero modes on β-$Bi_2Pd$ thin films." Science bulletin 62.12 (2017): 852-856.

[21] Li, Yufan, et al. "Observation of half-quantum flux in the unconventional superconductor β-$Bi_2Pd$." Science 366.6462 (2019): 238-241.

[22] Fang, Yuqiang, et al. "Discovery of superconductivity in 2M $WS_2$ with possible topological surface states." Advanced Materials 31.30 (2019): 1901942.

[23] Yuan, Yonghao, et al. "Evidence of anisotropic Majorana bound states in *2M*-$WS_2$." Nature Physics 15.10 (2019): 1046-1051.

[24] Chen, Dong-Yun, et al. "Superconducting properties in a candidate topological nodal line semimetal $SnTaS_2$ with a centrosymmetric crystal structure." Physical Review B 100.6 (2019): 064516.

[25] Cui, Zhihai, et al. "Type-II Dirac Semimetal State in a Superconductor Tantalum Carbide." Chinese Physics Letters 37.8 (2020): 087103.

[26] Shang, T., et al. "Superconductivity and topological aspects of the rock-salt carbides NbC and TaC." Physical Review B 101.21 (2020): 214518.

[27] Yan, Dayu, et al. "Superconductivity and Fermi Surface Nesting in the Candidate Dirac Semimetal NbC." Physical Review B 102.20 (2020): 205117.

[28] Klimczuk, Tomasz, and Robert Joseph Cava. "Carbon isotope effect in superconducting MgCNi$_3$." Physical Review B 70.21 (2004): 212514.

[29] Charles P. Poole et al., Superconductivity (Second Edition), Academic Press, 2007, pages: 113-142.

[30] Prozorov, Ruslan, and Vladimir G. Kogan. "Effective demagnetizing factors of diamagnetic samples of various shapes." Physical Review Applied 10.1 (2018): 014030.

[31] Tinkham M. Introduction to superconductivity [M]. Courier Corporation, 2004.

[32] Gruneisen, E. "The temperature dependence of the electrical resistance of pure metals." Annalen der Physik (Leipzig) 16 (1933): 530.

[33] Bid, Aveek, Achyut Bora, and A. K. Raychaudhuri. "Temperature dependence of the resistance of metallic nanowires of diameter $\geqslant$ 15 nm: Applicability of Bloch-Grüneisen theorem." Physical Review B 74.3 (2006): 035426.

[34] Ziman J M. Electrons and Phonons, Chap. 10[J]. 1960.

[35] Deutsch, M. "An accurate analytic representation for the Bloch-Gruneisen integral." Journal of Physics A: Mathematical and General 20.13 (1987): L811.

[36] Zeller, H. R. "Effect of lattice instabilities on superconducting and other properties in TaC$_x$." Physical Review B 5.5 (1972): 1813.

[37] Giorgi, A. L., et al. "Effect of composition on the superconducting transition temperature of tantalum carbide and niobium carbide." Physical Review 125.3 (1962): 837.

[38] McMillan, W. L. "Transition temperature of strong-coupled superconductors." Physical Review 167.2 (1968): 331.

[39] Amon, Alfred, et al. "Noncentrosymmetric superconductor BeAu." Physical Review B 97.1 (2018): 014501.

[40] Blackburn, Simon, et al. "Enhanced electron-phonon coupling near the lattice instability of superconducting NbC$_{1-x}$N$_x$ from density-functional calculations." Physical Review B 84.10 (2011): 104506.

[41] Taylor, O. J., A. Carrington, and J. A. Schlueter. "Specific-Heat Measurements of the Gap Structure of the Organic Superconductors κ−(ET)$_2$Cu [N(CN)$_2$]Br and κ−(ET)$_2$Cu(NCS)$_2$." Physical review letters 99.5 (2007): 057001.

[42] Xu, Xiaofeng, et al. "Evidence for two energy gaps and Fermi liquid behavior in the SrPt$_2$As$_2$ superconductor." Physical Review B 87.22 (2013): 224507.

[43] Niu, C. Q., et al. "Effect of selenium doping on the superconductivity of Nb$_2$Pd (S$_{1-x}$Se$_x$)$_5$." Physical Review B 88.10 (2013): 104507.

[44] Werthamer, N. R. "g. Helfand and PC Hohenberg." Phys. Rev 147 (1966): 295.

[45] Wei, Wensen, et al. "Rh$_2$Mo$_3$N: Noncentrosymmetric s-wave superconductor." Physical Review B 94.10 (2016): 104503.

[46] Yuan, H. Q., et al. "Nearly isotropic superconductivity in (Ba, K)Fe$_2$As$_2$." Nature 457.7229 (2009): 565-568.

[47] Khim, Seunghyun, et al. "Pauli-limiting effects in the upper critical fields of a clean LiFeAs single crystal." Physical Review B 84.10 (2011): 104502.

[48] Carnicom, Elizabeth M., et al. "TaRh$_2$B$_2$ and NbRh$_2$B$_2$: Superconductors with a chiral noncentrosymmetric crystal structure." Science advances 4.5 (2018): eaar7969.

[49] Singh, Yogesh, et al. "Multigap superconductivity and Shubnikov–de Haas oscillations in single crystals of the layered boride OsB$_2$." Physical Review B 82.14 (2010): 144532.

[50] Brandt, Ernst Helmut. "Properties of the ideal Ginzburg-Landau vortex lattice." Physical Review B 68.5 (2003): 054506.

[51] Mayoh, D. A., et al. "Superconducting and normal-state properties of the noncentrosymmetric superconductor Re$_6$Zr." Physical Review B 96.6 (2017): 064521.

[52] Hor, Yew San, et al. "Superconductivity in Cu$_x$Bi$_2$Se$_3$ and its implications for pairing in the undoped topological insulator." Physical review letters 104.5 (2010): 057001.

[53] Liu, Zhongheng, et al. "Superconductivity with topological surface state in Sr$_x$Bi$_2$Se$_3$." Journal of the American Chemical Society 137.33 (2015): 10512-10515.

[54] Qiu, Yunsheng, et al. "Time reversal symmetry breaking superconductivity in topological materials." arXiv preprint arXiv:1512.03519 (2015).

[55] Xu, Su-Yang, et al. "Momentum-space imaging of Cooper pairing in a half-Dirac-gas topological superconductor." Nature Physics 10.12 (2014): 943-950.

[56] Xu, Jin-Peng, et al. "Artificial topological superconductor by the proximity effect." Physical Review Letters 112.21 (2014): 217001.

[57] Guan, Syu-You, et al. "Superconducting topological surface states in the noncentrosymmetric bulk superconductor $PbTaSe_2$." Science advances 2.11 (2016): e1600894.

Table I. Atomic coordinates and equivalent isotropic thermal parameters of TaC.

| Site | Wyckoff | x | y | z | Occup[a] | $U_{eq}$[b] |
|---|---|---|---|---|---|---|
| Ta | 4a | 0 | 0 | 0 | 1 | 0.017(3) |
| C | 4b | 0 | 0.5 | 0 | 1 | 0.03(2) |

[a] *Occup*: Occupancy.
[b] $U_{eq}$: equivalent isotropic thermal parameter.

Table II. Measured and calculated superconducting parameters of TaC and NbC.

| Parameter | TaC | NbC[27] |
|---|---|---|
| $T_c$ | 10.3 K | 11.5 K |
| $\mu_0 H_{c1}(0)$ | 38.6 mT | 19.6 mT |
| $\mu_0 H_{c2}(0)$ | 0.3 T | 1.21 T |
| $\mu_0 H_c(0)$ | 136 mT | 107 mT |
| $l$ | 448 Å | 33 Å |
| $\xi_0$ | 543 Å | 1035 Å |
| $\xi_{GL}$ | 331 Å | 165 Å |
| $\lambda_{GL}$ | 595 Å | 1322 Å |
| $\kappa_{GL}$ | 1.8 | 8.01 |
| $\gamma_n$ | 2.21 mJ mol$^{-1}$ K$^{-2}$ | 3.01 mJ mol$^{-1}$ K$^{-2}$ |
| $\Delta C/\gamma_n T_c$ | 1.94 | 1.43 |
| $\mu_0 H_P$ | 19 T | 21 T |
| $\lambda_{ep}$ | 0.986 | 0.848 |
| $N(E_F)$ | 0.47 eV$^{-1}$ per f.u. | 0.69 eV$^{-1}$ per f.u. |
| $\theta_D$ | 213.5 K | 321.6 K |
| $r$ | 1.17 | 1.06 |
| $\Delta(0)$ | 1.72 meV | 1.85 meV |
| $2\Delta(0)/k_B T_c$ | 4.12 | 3.73 |

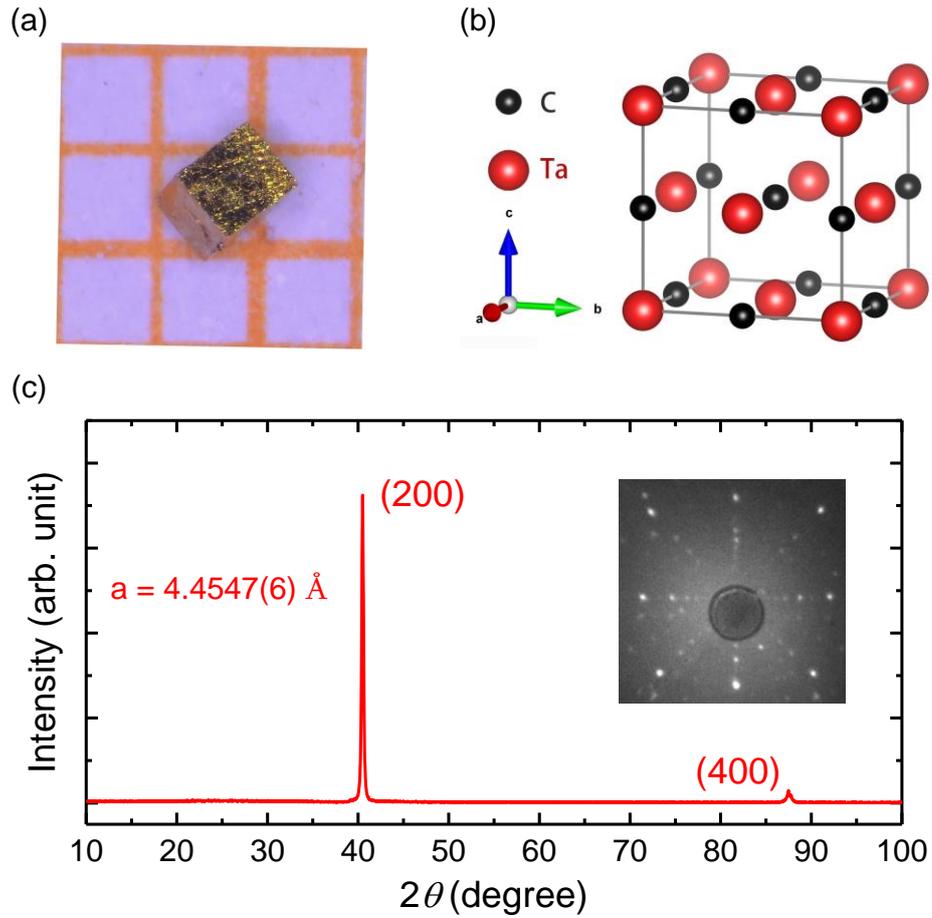

Figure 1 (a) Photograph of a typical TaC single crystal. (b) The schematic crystalline structure of TaC. (c) The XRD patterns of a flat surface of TaC single crystals. The inset shows the Laue diffraction pattern on the (100) surface.

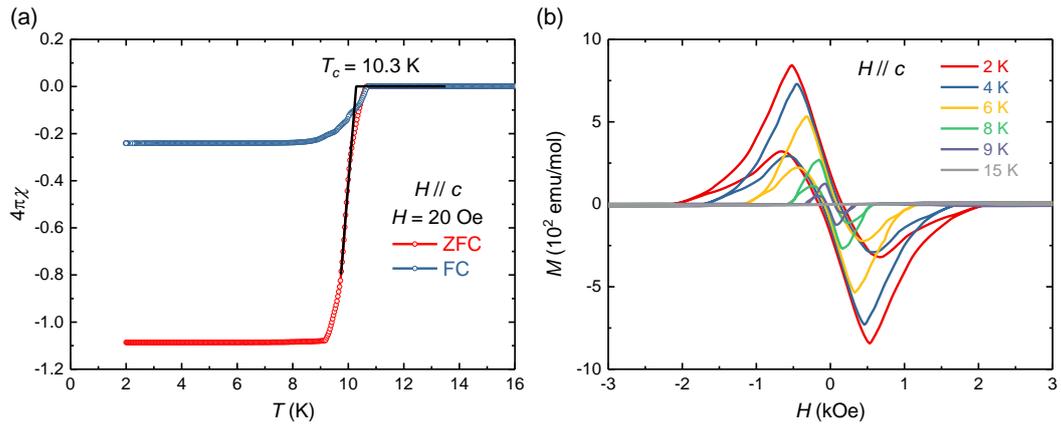

Figure 2 (a) Temperature dependence of $4\pi\chi$ for applied field $H$ = 20 Oe in ZFC and FC modes. (b) Isothermal *M-H* curves of TaC at various fixed temperatures.

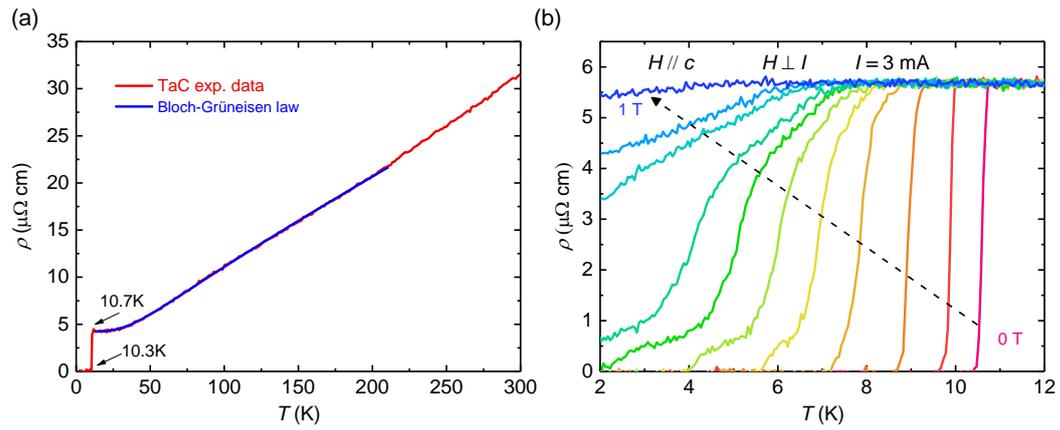

Figure 3 (a) Temperature dependence of longitudinal resistivity $\rho$ of TaC. (b) The resistivity at various magnetic fields.

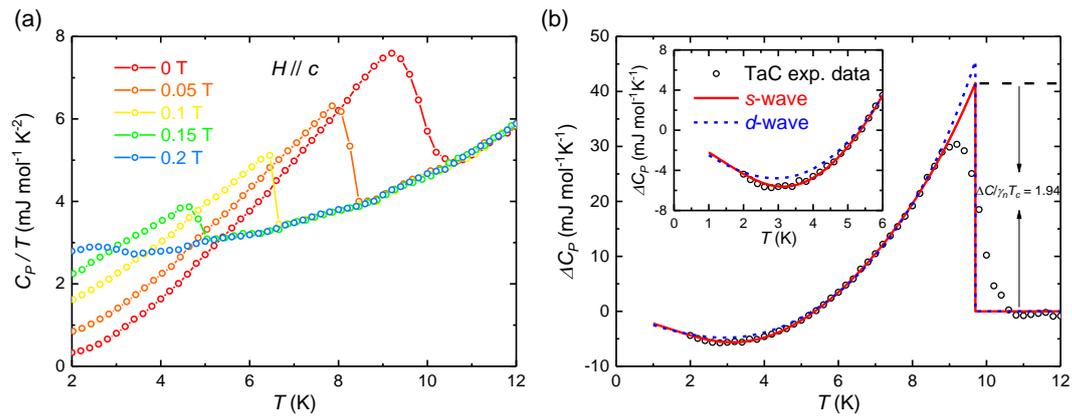

Figure 4 (a) Special heat $C_P/T$ versus $T$ curves at various applied magnetic fields. (b) Experiment data for ΔC vs T, plotted by fits using s-wave and d-wave gap function. The inset shows an enlarged image at low temperature.

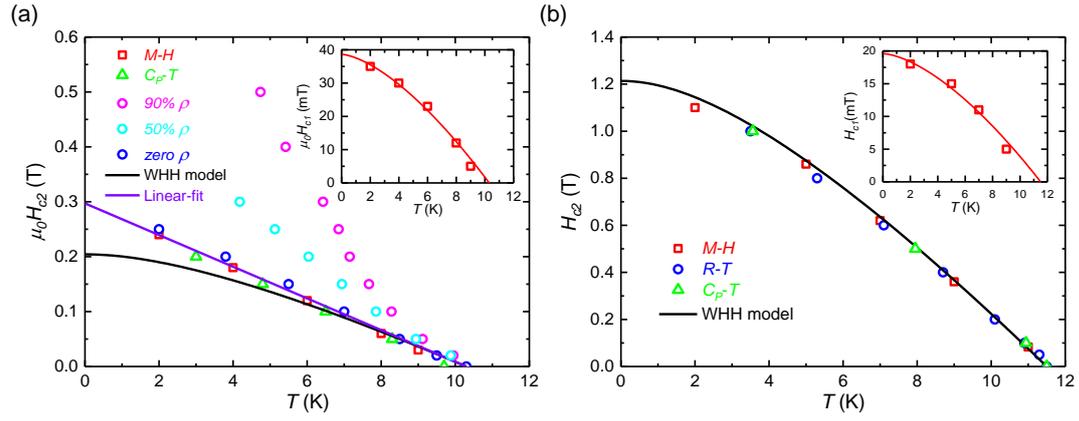

Figure 5 (a) Phase diagram of the upper critical field $H_{c2}$ of TaC versus $T$ with the fitting curves. The inset shows the lower critical field $H_{c1}$ versus $T$. (b) Phase diagram of NbC from Ref [27].